\documentclass[aps,pre,amsmath,floatfix,twocolumn,byrevtex,superscriptaddress]{revtex4}
\usepackage{epsfig}
\usepackage{graphicx}
\usepackage{amssymb}
\begin{document}
\title{One-dimensional cluster growth and branching gels in colloidal
systems with short-range depletion attraction and screened electrostatic repulsion}


\author{F. Sciortino} \affiliation{ {Dipartimento di Fisica and
  INFM-CRS-SOFT, Universit\`a di Roma {\em La Sapienza}, P.le A. Moro
  2, 00185 Roma, Italy} } \author{P. Tartaglia} \affiliation{
  {Dipartimento di Fisica and INFM-CRS-SMC, Universit\`a di Roma {\em
  La Sapienza}, P.le A. Moro 2, 00185 Roma, Italy} }
  \author{E. Zaccarelli} \affiliation{ {Dipartimento di Fisica and
  INFM-CRS-SOFT, Universit\`a di Roma {\em La Sapienza}, P.le A. Moro
  2, 00185 Roma, Italy} }
        
\begin{abstract}
We report extensive numerical simulations of a simple model for
charged colloidal particles in suspension with small non-adsorbing
polymers. The chosen effective one-component interaction potential is
composed of a short-range attractive part complemented by a Yukawa
repulsive tail. We focus on the case where the screening length is
comparable to the particle radius. Under these conditions, at low
temperature, particles locally cluster into quasi one-dimensional
aggregates which, via a branching mechanism, form a macroscopic
percolating gel structure. We discuss gel formation and contrast
it with the case of longer screening lengths, for which previous
studies have shown that arrest is driven by the approach to a Yukawa
glass of spherical clusters. We compare our results with recent
experimental work on charged colloidal suspensions [A. I. Campbell
{\it et al.} cond-mat/0412108, Phys. Rev. Lett. in press].
\end{abstract}

\maketitle

Recent years have witnessed a progressive interest in the role of the
inter-particle potential on controlling structure and dynamics of
colloidal dispersions.
Experiments
\cite{Seg01a,Din02a,Kegel03,Strad04,Sedg04,Bagl04,Kegel04,Bordi05,Sedg05,Bartlett04,Stiakpreprint},
theory\cite{Gro03a,ChenPRE,Liu05} and simulation
\cite{Sci04a,Sator03,Coni04,Imp04,Mos04a,Kumar05} studies have
provided evidence that when the hard-core repulsion is complemented
simultaneously by a short range attraction (of finite depth) and by a
screened electrostatic repulsion, particles tend to form aggregates,
whose shape and size is sensitively dependent on the balance between
attraction and repulsion\cite{Wud92a,Dee94a,Sear99,Mur02a,Mos04a}.
In some cases, the system shows an equilibrium cluster phase, where
particles associate and dissociate reversibly into
clusters\cite{Strad04,Sedg04,Bordi05}.  Interestingly enough, these
cluster phases appear not only in colloidal systems but also in
proteins solutions, in the limit of low salt
concentration\cite{Strad04,Bagl04,Sedg05}. Estimates of the ground state
configuration of isolated clusters of different size\cite{Mos04a}
suggest that, when the clusters diameter exceeds the screening length,
the shape of the aggregates crosses from spherical to linear. Evidence
has been reported that, for appropriate tuning of the external control
parameters, colloidal cluster phases progressively evolve toward an
arrested state\cite{Seg01a,Sedg04,Sedg05,Bartlett04}. It has been suggested, and
supported by numerical simulations, that, in the case of relatively
large screening length (i.e. the case of preferentially spherical
clusters), dynamic arrest may proceed via a glass transition
mechanism, where clusters, acting as {\it super}-particles interacting
via a renormalized Yukawa potential, become confined by the repulsions
created by their neighboring clusters\cite{Sci04a}. This mechanism is,
in all respects, identical to the glass transition of Yukawa
particles\cite{Lai97a,Lai98a,Bos98a,Wan99a} and leads, favored by the
intrinsic polydispersity of the clusters induced by the growth
process, to the realization of a Wigner glass. The simulation
study\cite{Sci04a} showed that the resulting arrested state is not
percolating, i.e. the arrest transition can not be interpreted in
terms of the formation of a bonded network of particles.

\begin{figure}
\hbox to\hsize{\epsfxsize=1.0\hsize\hfil\epsfbox{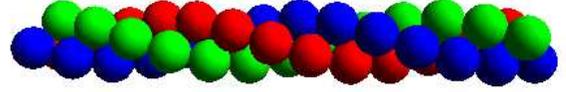}\hfil}
\caption{A pictorial view of the Bernal spiral.  Particles have been
differently colored to highlight the presence of three strands.  In
this geometry, each particle has exactly six nearest neighbors.}
\label{fig:spiral}
\end{figure}

A very recent experimental work\cite{Bartlett04} has reported evidence of
arrest via linear cluster growth followed by percolation, in a system
of charged colloidal particles. In the studied system, the short-range
attraction, induced via depletion mechanism, is complemented by an
electrostatic repulsion, with a Debye screening length $\xi$ estimated
of the order of $\xi/\sigma \approx 0.65$, where $\sigma$ indicates
the hard core diameter of the colloidal particle.  The quasi
one-dimensional clusters observed via confocal microscopy  are locally
characterized by a Bernal-spiral geometry\cite{Bernal}, the same structure
found as cluster ground state configuration for the case of
screening lengths smaller than $\sigma$ \cite{Mos04a}. The Bernal
spiral, shown in Fig.~\ref{fig:spiral}, is composed of face sharing
tetrahedra, in which each particle is connected to six neighbors.

In this work we numerically investigate the possibility that, when the
potential parameters are such that the Bernal spiral is the ground
state structure for isolated clusters, macroscopic gels can be formed
at large, but finite, attraction strength, via a mechanism of
branching favored by the small but finite thermal contributions.  We
explore the low packing fraction region for several values of the
attractive interaction strength, to highlight the collective effects
arising from cluster-cluster interactions and to assess under which
external conditions, ground state predictions are valid.
We carry our study along two routes.  In both cases, we study a
colloid-polymer mixture in the effective one-component description,
i.e. assuming that the polymer size is much smaller than that of the
colloids. In the first route, we control the attraction between
colloidal particles via a temperature scale. In the second route ---
designed to make direct contact with the experimental work reported in
Ref.\cite{Bartlett04} --- we study an isothermal system where the
repulsive part of the potential is fixed, while the attractive part of
the potential is varied according to the concentration of depletant,
to model the strength of the polymer-induced depletion interaction. We
will refer to these two set of simulations respectively as temperature
and polymer concentration route, naming them after the respective
relevant control parameters.

We study --- as a function of the packing fraction and of the
attraction strength --- the shape of the clusters (quantified via
their fractal dimension), the local geometry around each particle, the
inter-particle structure factor, the connectivity properties of the
system.  We complement the static picture with information on the
inter-particle bond lifetime and on the dynamics of self and collective
properties. We compare these quantities for the two routes, and show
that the two approaches provide a similar description of cluster
growth, percolation and gel formation.

\section{Simulation Details}

We study a system composed of $N=2500$ colloidal particles of diameter
$\sigma$ and mass $m$ in a cubic box of size $L$, as a function of the
packing fraction $\phi_c=\pi\rho\sigma^3/6$, where $\rho=N/L^3$ is the
number density, and of the temperature $T$. The particles interact
simultaneously via a short-range potential $V_{SR}$ and a
screened electrostatic repulsive interaction $V_{Y}$. The short-range
attraction is modeled for simplicity with the generalization to
$\alpha=18$ of the Lennard-Jones $2\alpha-\alpha$ potential, as proposed by
Vliegenthart {\em et al.}~\cite{Vli99})
\begin{equation} 
V_{SR}(r)=4 \epsilon  \bigg[ \bigg(\frac{\sigma}{r}\bigg)^{2\alpha}- 
\bigg(\frac {\sigma}{r}\bigg)^{\alpha}\bigg], 
\label{eq:potsr}
\end{equation}  
where $\epsilon$ is the depth of the potential. The parameters
$\sigma$ and $\epsilon$ are chosen as units of length and energy
respectively. We also consider $k_B=1$.  For this choice of $\alpha$ the
width of the attraction range is roughly $0.2 \sigma$.  The phase
diagram of $V_{SR}(r)$ has been studied previously~\cite{Vli99} and
it is characterized by a rather flat gas-liquid coexistence line, with
a critical point located at $T_c^{SR}\simeq 0.43$ and
$\phi_c^{SR}\simeq 0.225$.

The repulsive interaction is modeled by a
Yukawa potential
\begin{equation}
V_{Y}(r)=A\; \frac{e^{-r/\xi}}{r/\xi}.
\label{eq:potyuk}
\end{equation}
characterized by an amplitude $A$ and a screening length $\xi$.  We
focus on the case $\xi=0.5 \sigma$ and $A=8 \epsilon$, for which the
minimum of the pair potential $V_{SR}+V_{Y}$ is located at
$r=1.042\sigma$ corresponding to a potential energy
$E_{min}=-0.52\epsilon$.  With the present choice of $A$ and $\xi$,
the ground state configuration of an isolated cluster is known to be
the one-dimensional Bernal spiral, shown in
Fig.~\ref{fig:spiral}\cite{Mos04a}.  The Bernal spiral structure is
composed of face-sharing tetrahedrons, resulting in three twisting
strands of particles in such a way that each particle has six nearest
neighbors. In this geometry,  
for large $N$, the potential energy
per particle $E$ is $E=-1.36+2.10/N$ (always in units of
$\epsilon$). In the bulk of the spiral (far from side effects) $E$ is
about three times $E_{min}$, confirming that the attractive
interaction with the six neighbors provides most of the binding
energy.

In parallel, we also study the case in which the magnitude of the
attractive part changes to mimic the dependence of the depletion
interaction on polymer concentration $\phi_p$. As in
Ref.~\cite{Bartlett04}, we choose $\epsilon/k_BT= -14 \phi_p$, i.e.
the attraction strength is assumed to depend linearly on the fraction
of free volume occupied by polymers $\phi_p$.  In this case, as in
experiments, $T$ is kept constant to $k_B T =1$. To study a model as
close as possible to the experimental work of Ref.~\cite{Bartlett04},
we select $\xi=0.65 \sigma$, $A=10\epsilon$\cite{notabartlettpot}, and
$\alpha=10$.  For this value of $\alpha$, $r_{min}=1.07 \sigma$, in
agreement with the position of the maximum in the radial distribution
function $g(r)$, as extracted from data reported in
Ref.~\cite{Bartlett04}.

\begin{figure}
\hbox to\hsize{\epsfxsize=1.0\hsize\hfil\epsfbox{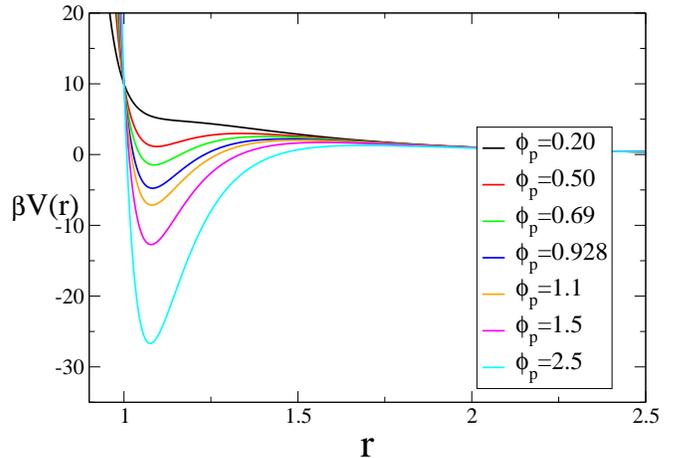}\hfil}
\caption{Interaction potential  $\beta V(r) \equiv \beta[V_{SR}(r)+V_{Y}(r)]$ for different values of the 
polymer concentration $\phi_p$.   Here $A=10 \epsilon$, $\xi=0.65 \sigma$, $\alpha=10$ and 
$\beta \epsilon= - 14.0 \phi_p$\protect\cite{Bartlett04}.
}
\label{fig:pot}
\end{figure}
The dependence of the potential shape with $\phi_p$ is shown in
Fig.\ref{fig:pot}.  Note that $V(r)$ changes from monotonically
repulsive to repulsive with a {\it local} minimum (with
$V(r_{min})>0)$. Finally, for $\phi_p>0.50$, $V(r)$ develops an
attractive global minimum followed by a repulsive tail.

In the rest of the present work, we will name $T$-route and $\phi_p$
route the two parallel set of simulations. The short-range nature of
$V_{SR}$ favors a very effective way to define pairs of bonded
particles. Indeed, the resulting potential $V(r)=V_{SR}+V_{Y}$ has a
well defined maximum located approximatively where the short range
attraction becomes negligible.  In the following we will consider
bonded (or nearest neighbors) all pairs of particles whose relative
distance $d<r_{max}$. In the $T$-route case, we fix
$r_{max}=1.28\sigma$, while in the $\phi_p$ case the bond distance
between two particles can be conveniently defined as $1.4 \sigma$.

\begin{figure}
\hbox to\hsize{\epsfxsize=1.0\hsize\hfil\epsfbox{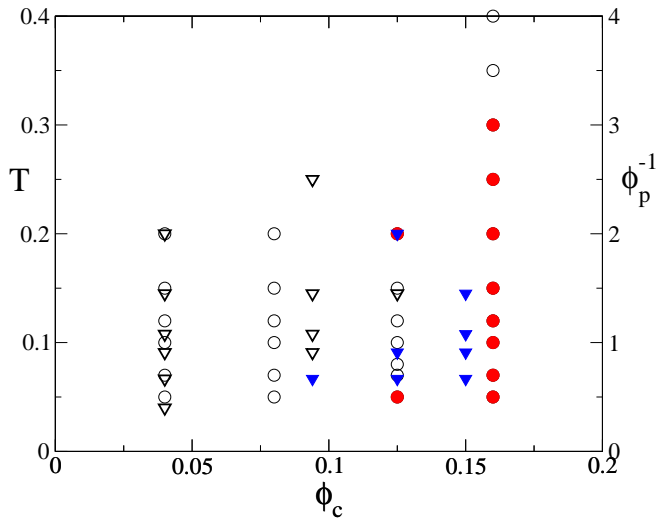}\hfil}
\caption{State points studied in this work, in the $T-\phi_c$ and $\phi_p-\phi_c$ planes, 
respectively for the $T$ (circles) and $\phi_p$ (triangles)   routes.
Full symbols indicate state points where the equilibrium structure presents a spanning network of
bonded particles. }
\label{fig:statepoints}
\end{figure}

In all simulations, time is measured in units of $\sqrt{m \sigma^2 /
\epsilon}$.  For numerical reasons, the repulsive potential is cut at
$r_c=8 \xi$, such that $V_Y(r_c) \approx
4.2~10^{-5} A$.  All simulated state points are shown in
Fig.\ref{fig:statepoints}. Equilibration is achieved with Newtonian
dynamics, followed by a Brownian dynamics simulation, based on the
scheme of Ref.\cite{All89}, to produce equilibrium trajectories. In the
case of Newtonian dynamics, the equation of motion have been
integrated with a time step of $\Delta t=0.02$. In the case of
Brownian dynamics, $\Delta t= 0.05$ with a bare diffusion coefficient
$D_o= 0.005 $. Equilibration runs required, at the slowest states,
more than $10^9$ integration time steps, corresponding to about three
months of computer time on a 1.6 GHz Pentium processor.

\section{Equilibration}
\label{sec:equil}

\begin{figure}
\hbox to\hsize{\epsfxsize=1.0\hsize\hfil\epsfbox{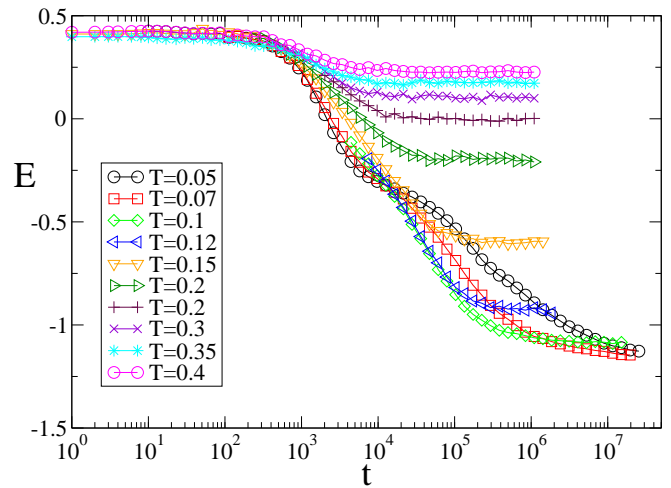}\hfil}
\caption{Time dependence of the potential energy 
following a quench starting from high temperature ($T=1.0$) 
for $\phi_c=0.16$, for the $T$-route case.}
\label{fig:epotR0.16}
\end{figure}

Simulations are started from high $T$ (or correspondingly
$\phi_p=0$) equilibrium configurations and quenched to the selected
final state. During equilibration, a Berendsen thermostat with a time
constant of $10$ is active, to dissipate the energy released in the
clustering process. Following the quench, the time evolution of the
potential energy $E$ shows a significant drop. Equilibration becomes
slower the lower the final $T$ or the larger $\phi_p$.  It also slows
down on lowering the colloid packing fraction $\phi_c$.  The evolution
of $E$ following a quench is shown in Fig.~\ref{fig:epotR0.16} for the
case $\phi_c=0.16$.  Around $T \lesssim 0.07$, equilibration cannot be achieved
within the simulation time and dynamic arrest takes place. In these
conditions, extremely slow (logarithmic in time) drift of $E$ is still
present at long times.  To provide evidence that equilibrium is
reached during the Newtonian simulation we check that $E$ is
independent on the previous history and that clusters reversibly
breaks and reform on changing $T$ or $\phi_c$.  Similar results are
obtained following the $\phi_p$-path.

The equilibration process is characterised by the progressive
formation of bonds between particles and the corresponding growth of
particle's aggregates, named clusters. 

\begin{figure}
\hbox to\hsize{\epsfxsize=1.0\hsize\hfil\epsfbox{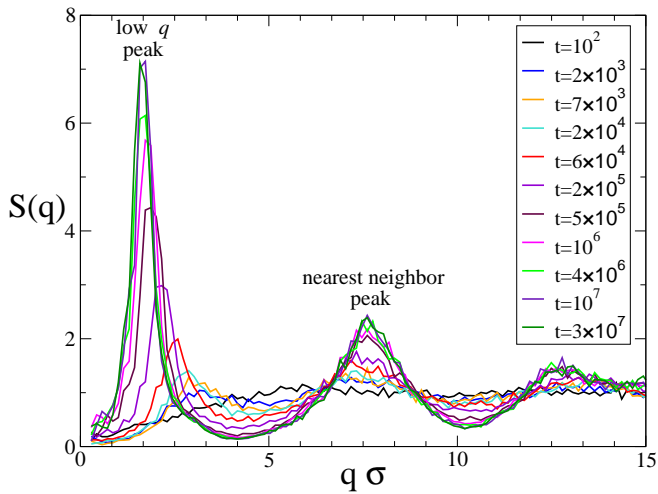}\hfil}
\caption{
Evolution of the static structure factor $S(q)$ during equilibration at  $\phi_c=0.125$ and $T=0.08$. }
\label{fig:sqvstime}
\end{figure}

A quantification of the evolution of the structure of the system
during equilibration can be provided by the structure factor $S(q)$,
defined as
\begin{equation}
S(q) = <\frac{1}{N} \sum_{i,j} e^{-i \vec q (\vec r_i -\vec r_j)}>
\end{equation}
where $\vec r_i$ indicates the coordinates of particle $i$.  The
$S(q)$ evolution, shown in Fig.~\ref{fig:sqvstime}, is
reminiscent of the initial stages of spinodal decomposition, showing a low $q$ peak which
grows in amplitude and moves to smaller and smaller $q$ vectors.
While in spinodal decomposition, the coarsening process proceeds
endless, in the present case the evolution of the small $q$ peak stops
when equilibrium is reached. The presence of the low $q$-peak in $S(q)$, at a
finite wavevector, highlights the presence of an additional
characteristic length scale in the system, discussed in more details
in the next section.

\begin{figure}
\hbox to\hsize{\epsfxsize=1.0\hsize\hfil\epsfbox{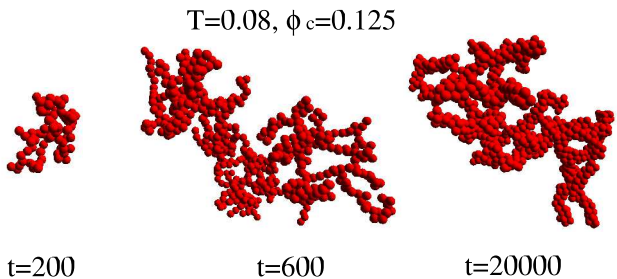}\hfil}
\caption{
Snapshots of the largest cluster at three different times during the equilibration process. 
Here $\phi_c=0.125$ and $T=0.08$. 
The cluster size  is 72, 605 and 908, respectively at $t=200$, $t=600$ and $t=20000$.}
\label{fig:timeevolution}
\end{figure}

Fig.~\ref{fig:timeevolution} shows the evolution of the shape of the
largest cluster for the case $\phi_c=0.125$ and $T=0.08$, one of the
cases in which the average cluster size grows monotonically in time.  It is interesting
to observe that at short times, the shape of the larger cluster is
rather ramified, the potential energy is still large and locally the
structure is still very different from the six-coordinated ground
state structure. Cluster arms are essentially composed by particles
arranged along lines. At longer times, the cluster arms get thicker
and thicker, and the local configuration approaches the characteristic
one of the Bernal spiral, even if some parts of the original branching
points persist in the final structure favoring the formation of a gel
network.  The evolution of the shape, complemented with the time
dependence of $E$, suggests that at large attraction
strengths (low $T$ or large $\phi_p$), the equilibration process can
be conceptually separated into two parts: an initial relaxation which
is closely reminiscent of the one which would take place if the
potential was purely attractive, followed by a second rearrangement
which sets in only after the coordination number has become
significant. At this point the competition of the long range
repulsion enters into play, forcing thereby the system to rearrange
into the expected local configuration. 
This competition results also in a non-monotonic evolution, during the equilibration, of the mean cluster size, at some state points.

\section{Equilibrium Properties: Statics}

\subsection{Potential Energy} 

\begin{figure}
\hbox to\hsize{\epsfxsize=1.0\hsize\hfil\epsfbox{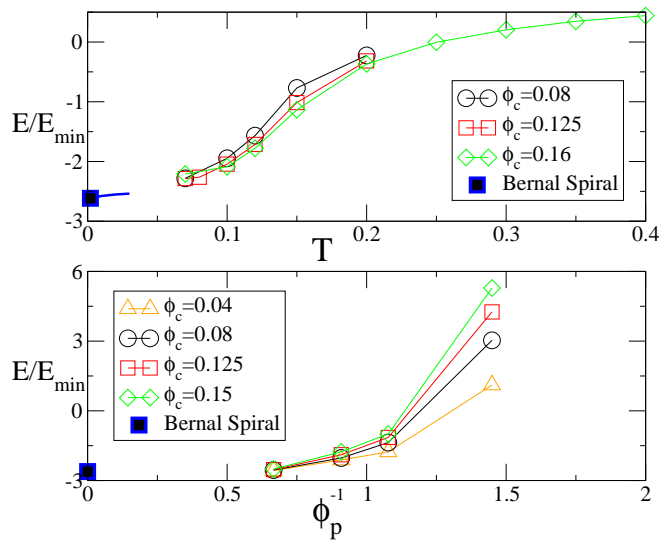}\hfil}
\caption{
$T$ (upper panel) and $\phi_p$ (lower panel) dependence of the normalized
potential energy per particle $E/E_{min}$ at different
$\phi_c$ values.   $E_{min}=-0.52 \epsilon$ in the $T$-route case, while it
depends on $\phi_p$  as $E_{min}=-14 \phi_p k T+V_Y(r_{min})$, with $r_{min}=1.07$ in the
$\phi_p$ case.  The corresponding  value for the Bernal spiral configuration is also reported.}
\label{fig:epotvsT}
\end{figure}

The upper panel of Fig.~\ref{fig:epotvsT} shows the $T$ dependence of
$E/E_{min}$ at the studied values of $\phi_c$. Around $T\approx 0.2$,
$E$ becomes negative, suggesting that the short-range attractive
interaction becomes relevant.  For lower $T$, $0.1<T<0.2$, $E$ drops
significantly, quickly reaching below $T=0.1$ a value compatible with
the ground state Bernal spiral configuration (also shown), once the
vibrational components are properly accounted for. A similar behaviour
is observed for the $\phi_p$ dependendence, shown in the lower panel
of Fig.~\ref{fig:epotvsT}.   In the studied $\phi_c$-range, the  $\phi_c$ dependence of $E$
is rather weak, especially for large attraction strengths.

\subsection{Cluster Size Distribution}

In this section  we examine the cluster size distribution, as it evolves with
$\phi_c$ and $T$. Standard algorithms are used to partition particles into clusters of
size $s$ and to evaluate the cluster size distribution $n_s$ and its
moments. The first moment of the cluster size distribution
\begin{equation}
<s>=\frac{\sum_s n_s s}{\sum_s n_s} = N/N_s
\end{equation}
is  connected to the inverse of the number of clusters $N_s$,  
while the second moment $<s_2>$ provides 
a representative measure of the average cluster size
\begin{equation}
<s_2> \equiv \frac{\sum_s s^2 n_s }{\sum_s s n_s}
\label{eq:s2}
\end{equation}

We also examine the connectivity properties of the equilibrium
configurations.  Configurations are considered percolating when,
accounting for periodic boundary conditions, an infinite cluster is
present.  The boundary between a percolating and a non-percolating
state point has been defined by the probability of observing infinite
clusters in 50$\%$ of the configurations.  To provide an estimate of
the percolation locus we report in Fig.~\ref{fig:statepoints} the
state points which are percolating.  We note that, at this level,
percolation is a geometric measure and it does not provide any
information on the lifetime of the percolating cluster.   Indeed, at $\phi_c = 0.125$,
percolation is present both at high $T$, where we observe geometric percolation of clusters with  bonds of very short life-time, and at low $T$ where the particles are connected by energetic bonds of very long life time, as discussed below. 
The competition between geometric and energetic percolation results in a
intermediate temperature window where the system does not percolate, i.e. in a re-entrant percolation locus.

\begin{figure}
\hbox to\hsize{\epsfxsize=0.89\hsize\hfil\epsfbox{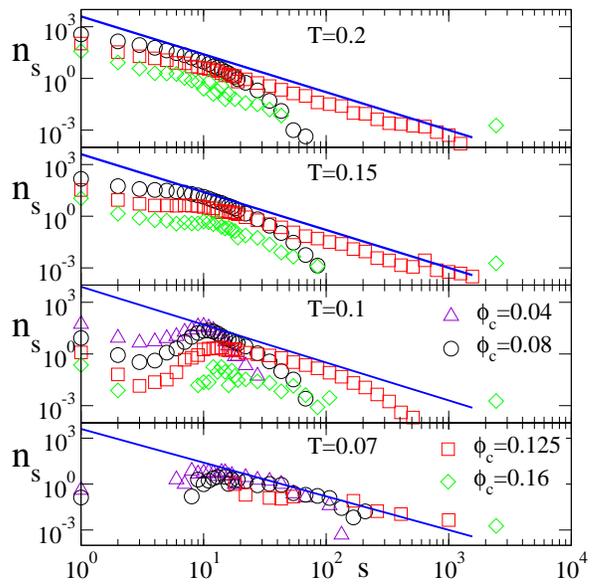}\hfil}
\vspace{0.10cm}
\caption{Cluster size distribution  $n_s$ at several $T$. In each panel, 
the full line represents the function $n_s \sim s^{-2.2}$.}
\label{fig:nsT}
\end{figure}

The cluster size distribution $n(s)$ is shown in
Fig.~\ref{fig:nsT}. At $T=0.2$ (where $E \approx 0$ and hence no significant bonding exists)
upon increasing $\phi_c$, the distribution progressively develops a
power-law dependence with an exponent $\tau$, consistent with the
random percolation value $\tau \approx
-2.2$\cite{Tor02book,Sta92book}. Percolation is reached when 
$0.125< \phi< 0.16$.
At slightly lower $T$, i.e. $T=0.15$, the picture remains
qualitatively similar, except for a hint of non-monotonic behavior,
around $s \approx 10-20$. On further lowering
$T$, the number of  clusters of  size $s \lesssim10$ drops significantly, to
eventually disappear at  $T=0.07$. These results are observed at all studied
densities.  

To frame the results presented above, we recall information previously
obtained in the study of the ground state energy of isolated cluster
of different size\cite{Mos04a}.  For a cluster size $s\lesssim 10$,
the addition of a monomer to an existing cluster lowers the energy per
particle, since the gain associated to the formation of an additional
attractive short-range bond is not yet compensated by the increased
number of repulsive interactions.  However, when clusters have grown
sufficiently, for $s \gtrsim10-20$, the energy driving force for
growing is reduced, since the energy per particle does not
significantly depend any longer on the cluster size\cite{Mos04a}.
Isolated clusters results carry on to the
interacting clusters case since the relatively small screening length
does not produce a significant cluster-cluster interaction. Indeed,
the effective cluster-cluster potential will be characterized, to a
first approximation, by the same $\xi$\cite{Sci04a}, which is short as
compared to the distance between clusters.

Around $T\approx 0.2$, $E(T=0.2) \approx 0$ and hence  energy can not be the driving force for clustering. Thus,  it is not a surprise
that close to percolation $n(s) \sim s^{-\tau}$ with $\tau$ consistent with the
random percolation value\cite{Tor02book,Sta92book}. The
disappearance of clusters of size $s \lesssim 10$, which starts to be
visible for $T \leqslant 0.1$, signals the progressive role of energy
in controlling clustering. At the lowest investigated $T$, energy has
taken over and all clusters are formed by energetically convenient
configurations.  In this respect, we can think of the low $T$ system
as a fluid composed of super-aggregates, providing an effective
re-normalization of the concept of "monomer" in the fluid. The small
cluster-cluster interaction energy  may favor a re-establishment of
the random percolation geometries and characteristic exponents, 
as discussed in the following.

\begin{figure}
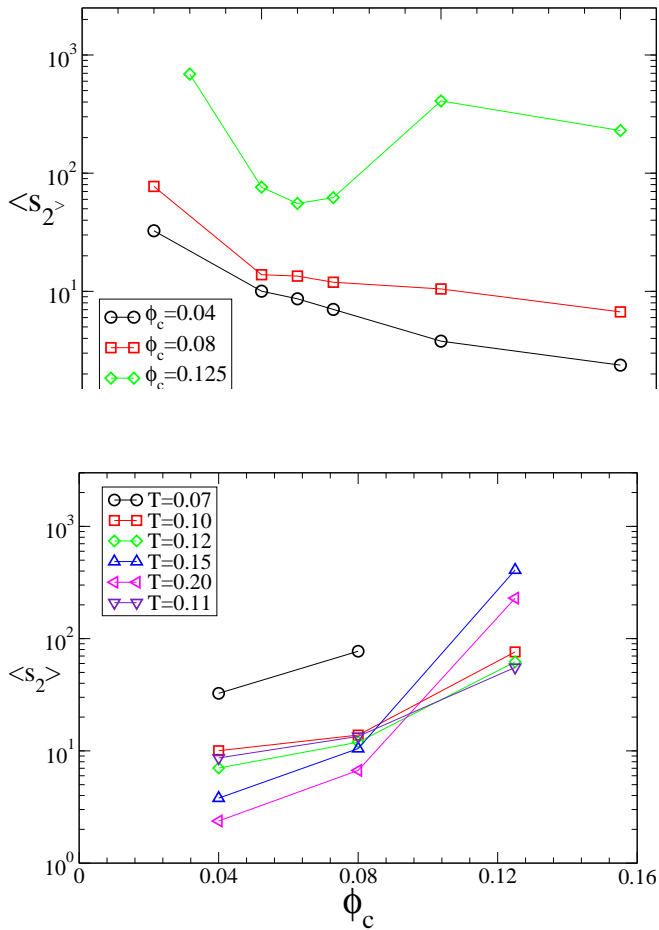

\hbox to\hsize{\epsfxsize=1.0\hsize\hfil\epsfbox{s2vsT-senzaperc.eps}\hfil}
\hbox to\hsize{\epsfxsize=1.0\hsize\hfil\epsfbox{s2vsphi-senzaperc.eps}\hfil}
\caption{Temperature and $\phi_c$ dependence of the second moment 
of the cluster size distribution $<s_2>$, for the $T$-route case.}
\label{fig:mcs}
\end{figure}

Fig.~\ref{fig:mcs} shows the $T$ and $\phi_c$ dependence of the second
moment of the distribution, the average cluster size $<s_2>$, defined
in Eq.\ref{eq:s2}, for all non-percolating state points.  Apart from
$\phi=0.16$, where configurations are percolating already before the
physics of the short-range bonding sets in, percolation at small
packing fractions is not reached at all temperatures we are able to
equilibrate.  At $\phi=0.125$, a non-monotonic dependence of
$<s_2>(T)$ is observed, which we interpret as a crossover from the
"random" percolation observed at high $T$ to the bond-driven
percolation, which becomes dominant at low $T$. The $\phi_c$
dependence of $<s_2>$ is shown in the bottom panel. At all $T$, a
monotonic growth is observed.

\subsection{Cluster shape}

\begin{figure}
\hbox to\hsize{\epsfxsize=1.0\hsize\hfil\epsfbox{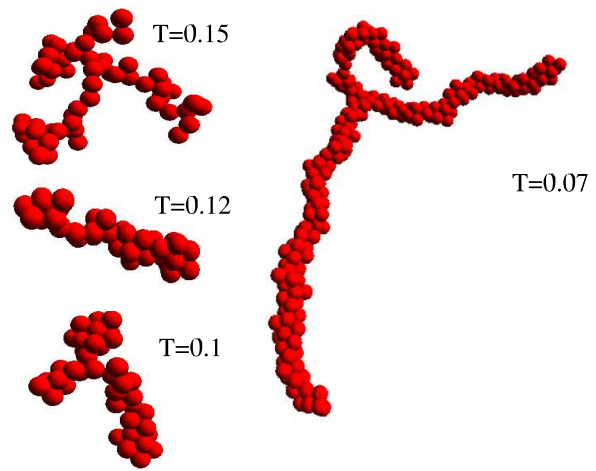}\hfil}
\caption{Typical largest  cluster  at 
$\phi_c=0.08$ for four different $T$ values: from
top left to bottom right
$T=0.15,0.12,0.1,0.07$.  }
\label{fig:largercluster0.08}
\end{figure}

\begin{figure}
\hbox to\hsize{\epsfxsize=1.0\hsize\hfil\epsfbox{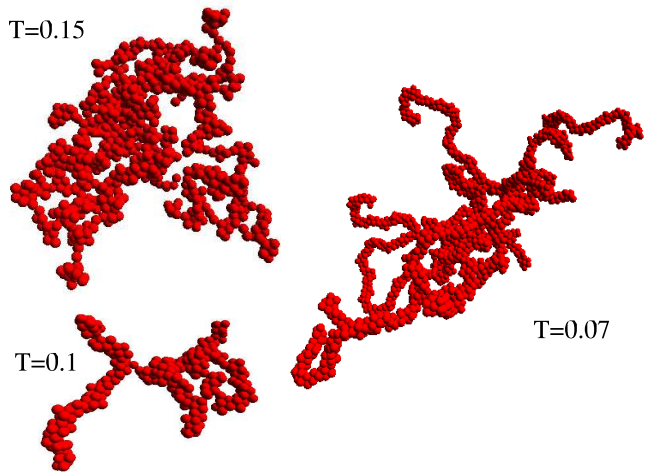}\hfil}
\caption{Typical largest  cluster  at $\phi_c=0.125$ for $T=0.15,0.1,0.07$.  
}
\label{fig:largercluster0.125}
\end{figure}

A pictorial description of the shape of the larger cluster observed in
a typical configuration at $\phi_c=0.08$ and $\phi_c=0.125$ for
different $T$ is shown in Figs.~\ref{fig:largercluster0.08} and
~\ref{fig:largercluster0.125}. In both cases, a progressive change of
shape of the largest cluster is observed on cooling. A close look to
the figures shows that on cooling particles become locally
tetrahedrally coordinated and that the loose high $T$ bonding
progressively crosses to a one dimensional arrangement of
tetrahedrons. At the lowest $T$, the clusters are composed by large
segments of Bernal spiral structures joined in branching points, the latter
providing the mechanism for network formation.
 
\begin{figure}
\hbox to\hsize{\epsfxsize=1.0\hsize\hfil\epsfbox{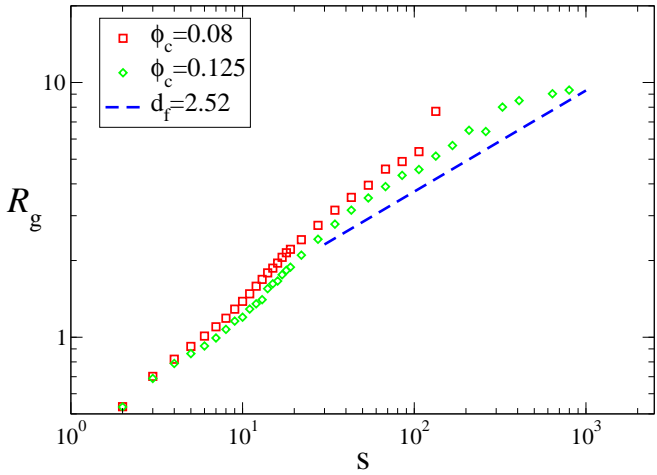}\hfil}
\caption{Size dependence of the cluster gyration radius at $T=0.15$ for two values of $\phi_c$.
The dashed line provides a reference slope for the random percolation $d_f$ value.}
\label{fig:gyration-highT}
\end{figure}

\begin{figure}
\hbox to\hsize{\epsfxsize=1.0\hsize\hfil\epsfbox{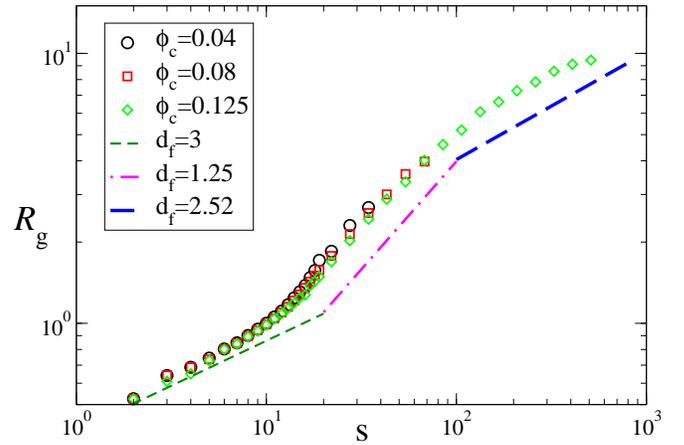}\hfil}
\caption{Size dependence of the cluster gyration radius at $T=0.1$ and several $\phi_c$. 
Lines provide reference slopes for different $d_f$ values.}
\label{fig:gyration-lowT}
\end{figure}

To quantify  the cluster shape 
we study the the cluster size dependence of the  cluster radius of gyration $R_g$, defined as 
\begin{equation}
R_g=\left[\frac {1}{N} \sum_{i=1}^{N}({\bf r}_i-{\bf R}_{CM})^2
\right]^{1/2}
\end{equation}
where ${\bf R}_{CM}$ are the center of mass coordinates.  For fractal
aggregates, $R_g \sim s^{1/d_f}$, where $d_f$ indicates the fractal
dimension.  The observed behavior of the clusters shape is very
different at high and low $T$.  Fig.\ref{fig:gyration-highT} shows
$R_g$ vs $s$ for two representative state points at $T=0.15$, close to
percolation.  The typical shape of the cluster at these two state
points is reported in Fig.~\ref{fig:largercluster0.08} and
Fig.~\ref{fig:largercluster0.125}.  When the cluster size is greater
than 20 monomers, the fractal dimension is consistent with the random
percolation value in three dimensions
($d_f=2.52$\cite{Tor02book,Sta92book}).  This value confirms that at
high $T$, as discussed previously, the energetic of the bonds is
negligible as compared to entropic effects and the cluster size grows
on increasing $\phi_c$, mostly due to the  increase in the average number of
particles with a relative distance less than $r_{max}$.  At low $T$,
an interesting phenomenon occurs, shown in
Fig.~\ref{fig:gyration-lowT}. The very small clusters ($s<10$) are
rather compact and $d_f \approx 3$ and indeed, in this size interval,
the energy per particle in the cluster decreases on increasing cluster
size\cite{Mos04a}.  For clusters with intermediate size $10 \lesssim s
\lesssim 100$, $d_f \approx 1.25$, supporting the preferential
one-dimensional nature of the elementary aggregation process, driven
by the repulsive part of the potential.  This $d_f$ value is observed
for all equilibrium cluster phases in which clusters of size $10 < s <
100$ are dominant, with a small trend toward smaller values for
smaller $T$ and $\phi_c$. This small $d_f$ value provides further
evidence that in this size interval growth is essentially uniaxial,
and that clusters of size 100 or less are essentially composed by
pieces of Bernal spirals joined by few branching points\cite{Mos04a}
(see Fig.~\ref{fig:largercluster0.08} and
Fig.~\ref{fig:largercluster0.125}).  For larger $s$ values, a
crossover toward $d_f \approx 2.52$ is observed. This crossover
suggests that for larger clusters the one-dimensional bundles have
branched a significant number of times, generating clusters whose
geometry is again controlled by random percolation features. Pieces of
Bernal spirals act as building blocks connected at branching
points in a random fashion.

\subsection{Structure Factor}

\begin{figure}
\hbox to\hsize{\epsfxsize=1.0\hsize\hfil\epsfbox{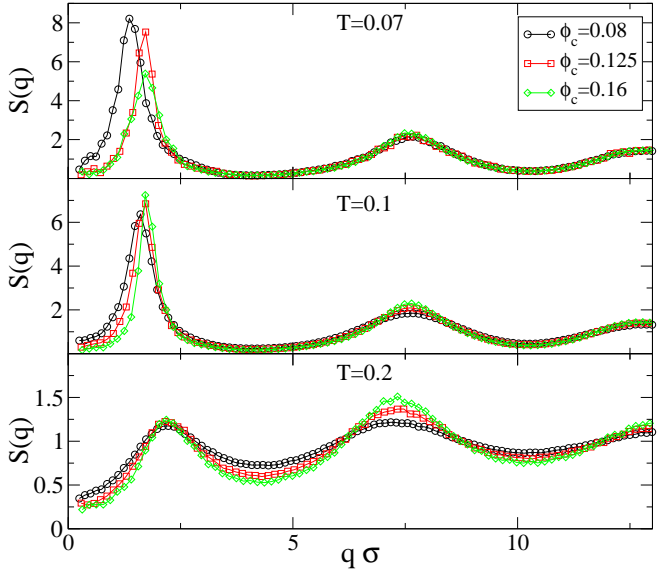}\hfil}
\caption{Wavevector $q$ dependence of $S(q)$ at three different  $T$  ($T=0.07, 0.1,
0.2$) for the $T$-route case.  For each $T$, data at three different $\phi_c$ are reported (
$\phi_c=0.08,
0.125, 0.16$). 
}
\label{fig:sqT}
\end{figure}

\begin{figure}
\hbox to\hsize{\epsfxsize=1.0\hsize\hfil\epsfbox{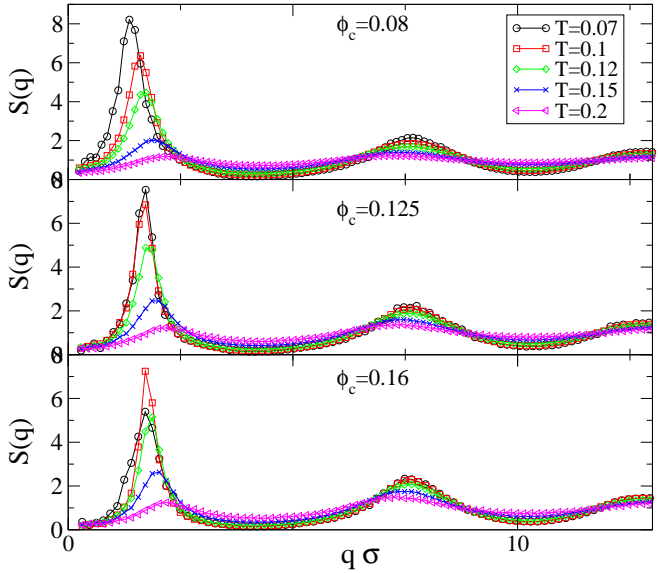}\hfil}
\caption{Wavevector $q$ dependence of $S(q)$ at three different  $\phi_c$   (
$\phi_c=0.08, 0.125, 0.16$) for the $T$-route case.  For each $\phi_c$, data at several different $T$ are reported.}
\label{fig:sqF}
\end{figure}

As discussed in Sec.~\ref{sec:equil}, the clustering process and the residual
repulsive interactions between different clusters produce an
additional low $q$ peak in $S(q)$, located well below  the
location of the nearest neighbor peak ($q \sigma \approx  2\pi$). 
Fig.~\ref{fig:sqT} and Fig.~\ref{fig:sqF} show respectively the $T$ and
$\phi_c$ dependence of $S(q)$, in equilibrium. 
Data refer to both percolating and non-percolating state points.
We observe no dependence of the position (either with $T$ or
$\phi_c$) of the nearest-neighbour peak, consistent with the presence of   
a deep minimum in the interaction potential, which
defines quite sharply the interparticle distance.  The amplitude of the nearest-neighbour 
peak grows 
on decreasing $T$ or increasing $\phi_c$. 
The location of the cluster-cluster peak  shows a weak $\phi_c$ dependence, 
 almost absent at $T=0.2$ (and
higher $T$), but which becomes more relevant at very low $T$.  We note that,
on isothermally increasing $\phi_c$, the location of the peak does not change
even when percolation is crossed. 
On the other hand, the $T$-dependence  is significant and the location of the peak
moves to smaller $q$ on decreasing $T$, suggesting the establishment of 
longer correlation lengths.  
The $T$ and $\phi_c$ trends are quite similar to those recently observed in
concentrated protein solutions at low ionic
strength\cite{Strad04}. In
particular, in that paper, the independence of the cluster peak
position on $\phi_c$ was interpreted as evidence of a linear dependence
of the equilibrium cluster size with $\phi_c$.  Indeed, if  clusters are assumed to be rather monodisperse in size and if  the
inverse of the peak position is assumed to be a measure of the
inter-cluster distance, the number cluster density has also to be
independent on $\phi_c$\cite{Strad04}.  It is worth stressing that, in one of the first
papers addressing the possibility of equilibrium cluster phases in
colloidal systems\cite{Gro03a,Groe04}, the same relation between equilibrium cluster size
and $\phi_c$ was presented, although its validity was limited to the
case of clusters of size significantly larger than the one observed
experimentally in Ref.\cite{Strad04}, hinting to a wider validity
of the relation suggested in Ref.~\cite{Strad04}. Here we note that the
independence of the $S(q)$ cluster peak position with $\phi_c$ holds
from very small $\phi_c$ up to values well beyond percolation, where an
interpretation in terms of finite clusters relative distance is
clearly not valid. 
In the present study (of non spherical clusters), we can access both $S(q)$ and the cluster size distribution.  We note that, as shown in    Fig.\ref{fig:nsT}, the cluster size distribution is not peaked around a typical value. Actually, the cluster size is significantly  non-monodisperse, expecially close to percolation. We also note that  neither $<s_1>$ nor $<s_2>$ (see Fig.~\ref{fig:mcs})
scale linearly with $\phi_c$, despite the constant position of the low $q$ peak in $S(q)$.


It would be relevant to understand how the parameters
$A$ and $\xi$ entering the potential (see Eqs.\ref{eq:potsr},\ref{eq:potyuk}), 
control the position of
the cluster peak and its $T$ and $\phi_c$ dependence.  In the case of
spherical clusters, it was possible to associate the peak position to
the average distance between clusters, since no percolation was
observed. This explanation is not fully satisfactory for
the present model, since, as can be seen in Fig.~\ref{fig:sqF} for the
case of $T=0.2$, the location of the peak is clearly the same both in
the non-percolating state $\phi=0.125$ and in the percolating state
$\phi=0.16$.  A better understanding of the quantities
controlling the peak position is requested. A first attempt in this direction has been recently 
presented \cite{Liu05}.

\subsection{Local Order} 
\begin{figure}
\hbox to\hsize{\epsfxsize=1.0\hsize\hfil\epsfbox{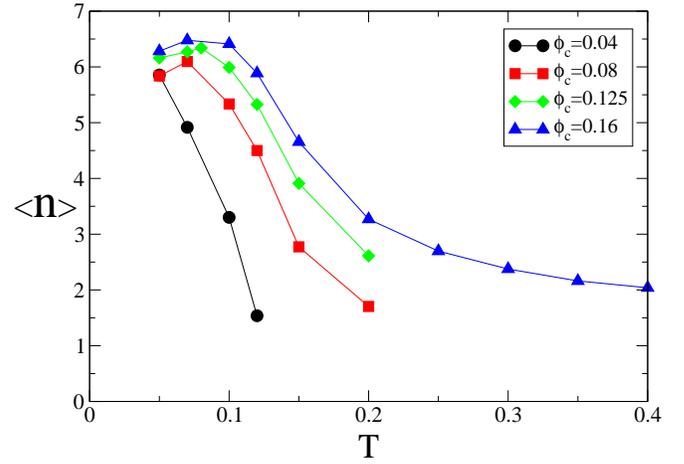}\hfil}
\caption{
Average number of neighbors $<n>$ as a function of $T$ for different $\phi_c$ values. 
Note that, for all  $\phi_c$, all curves approach 
the $<n>=6$ value characteristic of the geometry of the Bernal spiral.}
\label{fig:avrneig}
\end{figure}

A simple and useful indicator of local order is provided by the average number of
nearest neighbors $<n>$ and by  the associated distribution of
nearest neighbors $P(n)$,  which counts the fraction of particles surrounded by
$n$ neighbors within $r_{max}$.  As shown in Fig.\ref{fig:avrneig}, $<n>$
grows upon progressively lowering $T$, approaching, in a non monotonous way, a coordination
number of six.

\begin{figure}
\hbox to\hsize{\epsfxsize=1.0\hsize\hfil\epsfbox{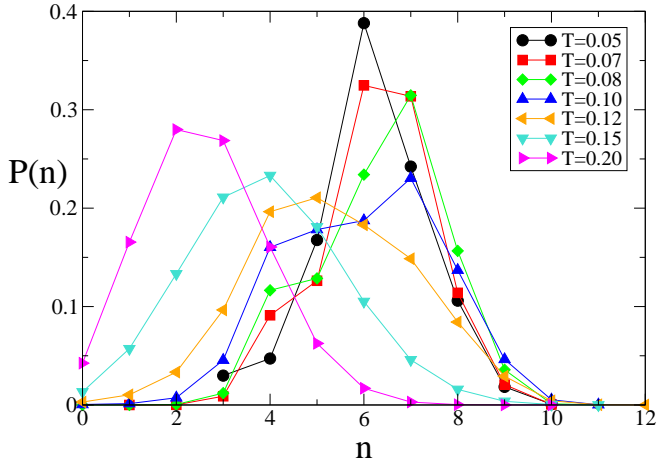}\hfil}
\caption{
Distribution 
of the number of neighbors $P(n)$ for several $T$ at $\phi_c=0.125$ (for the $T$-route case).}
\label{fig:neig}
\end{figure}

Fig.~\ref{fig:neig} shows the $T$ evolution of the distribution
$P(n)$.  Again, a clear preference for local geometries with about
six neighbors is displayed at low $T$ a condition which is hardly
observed in other materials in which particle-particle interaction is spherically symmetric. The
value $<n> =6$  is consistent with a local
geometry of face-sharing tetrahedra.

Another useful indicator of local order, which enables us to
effectively quantify the local structure, is provided by the
so-called local orientation order parameters $\bar q_{lm}(i)$ defined
as,
\begin{equation}
\bar q_{lm}(i) \equiv \frac{1}{N_{b_i}} 
\sum_{j=1}^{N_{b_i}}  Y_{lm}(\hat r_{i j})
\end{equation}
where $N_{b_i}$ is the set of bonded neighbors of a particle $i$.  The
unit vector $\hat r_{ij}$ specifies the orientation of the bond
between particles $i$ and $j$. In a given coordinate frame, the
orientation of the unit vector $\hat r_{ij}$ uniquely determines the
polar and azimuthal angles $\theta_{i j}$ and $\phi_{ij}$.  The
$Y_{lm}(\theta_{i j},\phi_{ij}) \equiv Y_{lm}(\hat r_{i j})$ are the
corresponding spherical harmonics.  Rotationally invariant local
properties can be constructed by appropriate combinations of the $\bar
q_{lm}(i)$. In particular, local order in crystalline solids, liquids
and colloidal gels, has been quantified focusing on
\begin{equation}
q_{l}(i) \equiv  \left[\frac{4 \pi}{2l+1} \sum_{m=-l}^{l} | 
\bar q_{lm}(i)|^2 \right]^{1/2}
\end{equation}
and 
\begin{equation}
\hat w_{l}(i) \equiv  w_l(i) / \left[ \sum_{m=-l}^{l} | 
\bar q_{lm}(i)|^2 \right]^{3/2}
\end{equation}
with 
\begin{equation}
w_{l}(i) \equiv  \!\!\!\!\!\!
\sum_{ \scriptsize \begin{array}{c} m_1,m_2,m_3, 
\\m_1\!+\!m_2\!+\!m_3\!=\!0\end{array}}\!\!\!\!\!\!    
\left( \begin{array}{ccc}
l & l & l \\
m_1 & m_2 & m_3 \end{array} \right) 
\bar q_{lm_1}(i)\bar q_{lm_2}(i)\bar q_{lm_3}(i).
\end{equation}

\begin{figure}
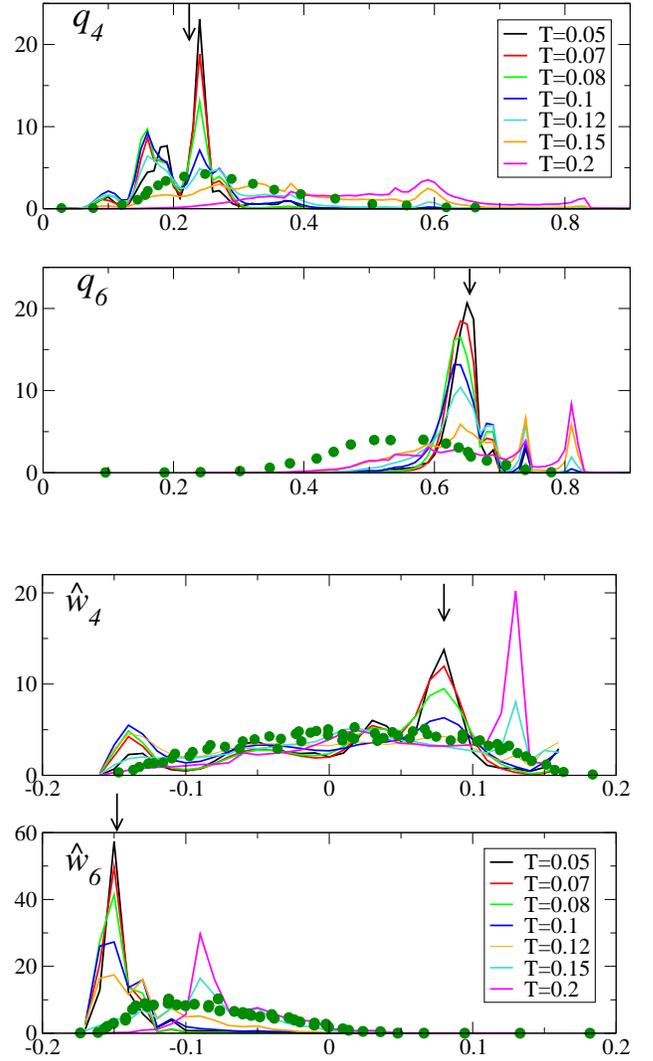

\hbox to\hsize{\epsfxsize=0.95\hsize\hfil\epsfbox{q.eps}\hfil}
\vspace{1.0cm}
\hbox to\hsize{\epsfxsize=0.95\hsize\hfil\epsfbox{w.eps}\hfil}
\caption{Temperature dependence of the rotational invariant distributions $P(q_i)$ (top) and $P(\hat w_i)$ (bottom) for $l=4$ and $l=6$
at $\phi=0.125$.  Arrows indicate the  ideal
Bernal spiral values.  In the ideal spiral, the local surrounding of all particles is identical and hence the rotational invariant distributions are delta functions.} 
\label{fig:ql}
\end{figure}

\begin{figure}
\hbox to\hsize{\epsfxsize=1.0\hsize\hfil\epsfbox{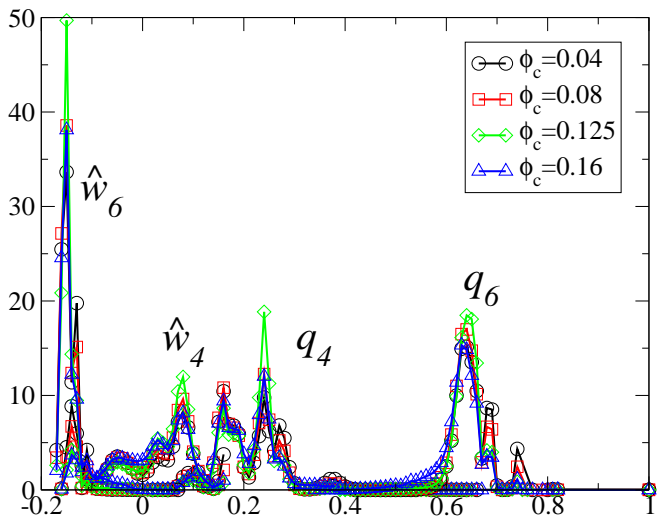}\hfil}
\caption{Packing fraction dependence of the rotational invariant distributions $P(q_i)$ and $P(\hat w_i)$ for $l=4$ and $l=6$ at $T=0.07$. Note that, at this low $T$, no $\phi_c$-dependence is present.}
\label{fig:distqwt007}
\end{figure}

The distributions of the $q_l$ and $\hat w_l$ parameters provide a
sensitive measure of the local environment and bond organization.  For
example, dimers are characterized by $q_l=1$, $\hat w_4=0.13$ and
$\hat w_6=-0.09$. A local tetrahedral order is characterized by large
negative values of $\hat w_6$, up to the value $-0.17$ for the
icosahedron \cite{Stein83}.  For the perfect Bernal spiral of
Fig.\ref{fig:spiral}, the orientational order parameters are determined as
$q_4=0.224$, $q_6=0.654$, $\hat w_4=0.08$ and $\hat w_6=-0.148$.
Fig.~\ref{fig:ql}
shows the $q_4$, $q_6$, $\hat w_4$ and $\hat w_6$ distributions and
how they evolve with decreasing temperature for $\phi=0.125$.  We note
that, upon cooling, the progressive presence of dimers and small
clusters disappears and the distributions evolve toward a limiting
form which appears to be specific of the Bernal spiral type of
cluster. At low $T$,  and in
particular below $T=0.1$, all distributions peak close to the characteristic values of the Bernal spiral.
The local orientation order parameters have been evaluated  in the confocal experimental
work of Ref.  \cite{Bartlett04}. There, it was shown that  the experimental data are consistent
with the Bernal geometry. In the analysis of the experimental data, the position of the particles in the perfect spiral geometry was subjected to some random displacements, to account for  thermal  fluctuations, possible intrinsic errors in the localization of the particles and polydispersity in size (and/or charge) in the samples. After this procedure,  the sharp peaks displayed in Figs.~\ref{fig:ql} and
\ref{fig:distqwt007} disappear and smooth distributions are obtained, which well compare with the
experimental data.  

Fig.~\ref{fig:distqwt007} shows that, at low $T$, the distributions appear to be insensitive
to $\phi_c$, in agreement with observations in \cite{Bartlett04} and supporting once more that the local structure around the majority of the particles  is similar to the ground state structure provided by the Bernal spiral.

\section{Dynamics and Gel Formation}

\begin{figure}
\hbox to\hsize{\epsfxsize=1.0\hsize\hfil\epsfbox{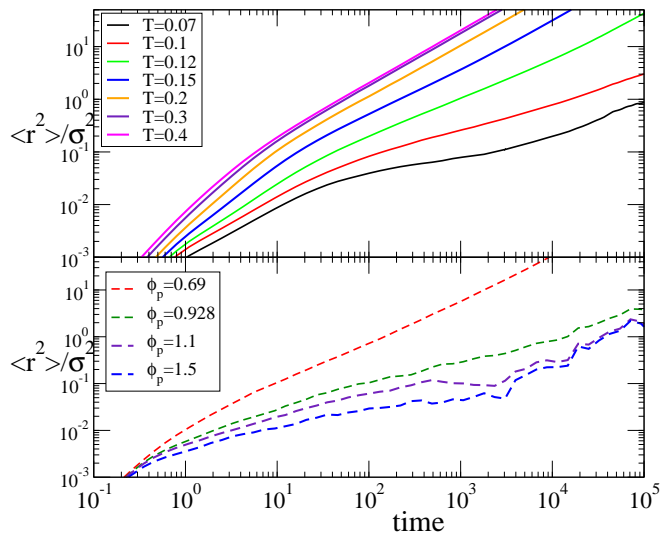}\hfil}
\caption{Averaged mean square displacement $<r^2>$  for the $T$-route (top) and the $\phi_p$-route (bottom)in log-log scale.  In the top panel, $\phi_c=0.16$, while in the bottom one $\phi_c=0.15$.}
\label{fig:r2R016}
\end{figure}

In this section, we present results for the particle dynamics as a
function of  $\phi_c$ and $T$ (in the $T$-route), or $\phi_p$ (in the $\phi_p$ route).
As for the equilibrium data shown in the previous section, dynamical quantities are evaluated from
trajectories generated according to a Brownian dynamics.  The mean
square displacement, $<r^2(t)>$, averaged over all particles and several
starting times is shown in Fig.~\ref{fig:r2R016} for one specific $\phi_c$ value  both for the $T$ and the $\phi_p$ routes. 

Beyond the ballistic region (which extends up to $<r^2(t)> \lesssim 10^{-3} \sigma^2$), particles enter in a diffusive regime,  composed of two
different processes. A short transient where the
bare self-diffusion coefficient $D_o$, set by the Brownian algorithm,
dominates and long-time region when particles feel the interparticle bonding.   
 At high $T$, in the latter regime, particles
diffuse almost freely, with a diffusion coefficient not very different
from the bare self-diffusion $D_o$ value.  Upon cooling, $<r^2(t)>$
progressively develops a plateau, more evident for $T\lesssim 0.1$,
which reaches the value $\approx 4 \cdot 10^{-2}$.  If we look at the
$\phi_p$-data, we observe a very similar behaviour, with a very
similar plateau which develops for $\phi_p \gtrsim 0.9$.  These results signal
that particles become tightly caged, with a localization length not
very different from the one observed in the case of dynamic arrest in glass forming systems, although in the present case caging is much less resolved.
Increasing the attraction strength, the long time limit
of $<r^2(t)>$ remains proportional to $t$, but with a smaller and smaller
coefficient.

\begin{figure}
\hbox to\hsize{\epsfxsize=1.0\hsize\hfil\epsfbox{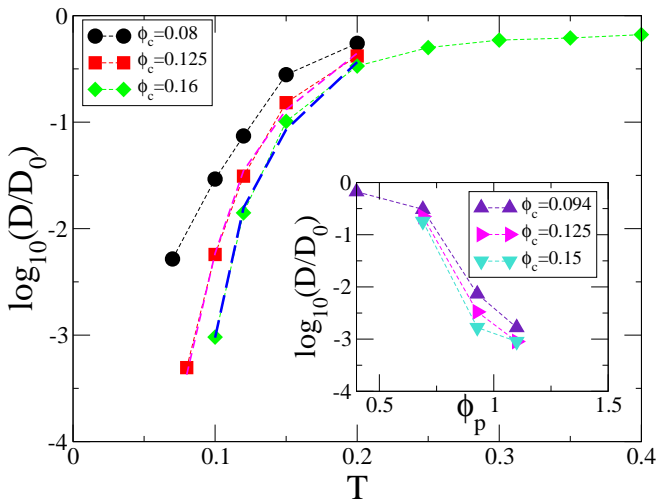}\hfil}
\caption{
Temperature dependence of the normalized diffusion 
coefficient $D/D_o$, for different $\phi_c$ values. 
The short and long dashed lines represent 
power law fits with exponent 
$\gamma_D=2.15$ and
$\gamma_D=2.37$ 
and dynamic critical temperatures $T_d=0.084$
and $T_d=0.091$  
respectively for $\phi_c=0.125$ and $\phi_c=0.16$. 
The inset shows the corresponding quantity for the $\phi_p$-route.}
\label{fig:diff}
\end{figure}

A global view of the $T$ and $\phi_c$ dependence of the slow dynamics is
shown in Fig.~\ref{fig:diff}, where the long time limit of
$<r^2(t)>/6t$, i.e. the self diffusion coefficient $D$, is
reported. While at high $T$ the diffusion coefficient approaches the
bare self-diffusion coefficient, on cooling, in the same $T$ interval
in which a substantial bonding takes place, $D$ drops several order of
magnitudes, signaling a significant slowing down of the dynamics and
the approach to a dynamically arrest state.  The same behaviour is
evident for the $\phi_p$ route, where $D$ approaches a very small value for $\phi_p
\geq 1.1$.  

It is interesting to note that for  $\phi_c=0.125$ and $\phi_c=0.16$, the $T$ dependence of $D$ is compatible with a power law, with exponent $\gamma_D \approx 2.2$, not very
different from the typical values of $\gamma_D$ predicted by MCT for simple liquids.  
The case of  $\phi_c=0.16$ is particularly interesting, since at all $T$, 
the instantaneous configuration of the system is percolating, 
providing a clear example of the difference between percolation and dynamic arrest.  
Vanishing of $D$ is observed only at very low $T$, well below percolation.  
It is tempting to state that, when the cluster-cluster interaction is weak 
as in the present case, dynamic arrest always requires the establishment of a percolating network of attractive bonds, though 
this is not a sufficient condition since the bond lifetime should
be significantly long.  When repulsive cluster-cluster interactions 
are not negligible, arrest at low $\phi_c$ can be generated in 
the absence of percolation\cite{Sci04a} via a Yukawa glass mechanism.

Another important quantity to characterize dynamic
arrest (particularly relevant for attraction-driven slowing down
\cite{Zac04JPCM}), is the bond correlation function
$\phi_B(t)$,  defined as
\begin{equation}
\phi_B(t)=\langle \sum_{i<j} n_{ij}(t)n_{ij}(0)\rangle /
[N_B(0)].
\end{equation} 
Here $n_{ij}(t)$ is 1 if two particles are bonded and 0 otherwise,
while $N_B(0) \equiv \langle \sum_{i<j}n_{ij}(0) \rangle $ is the
number of bonds at $t=0$. The average is
taken over several different starting times. $\phi_B$ counts which fraction of bonds
found at time $t=0$ are still present after time $t$,
independently from any breaking-reforming intermediate process.

\begin{figure}
\hbox to\hsize{\epsfxsize=1.0\hsize\hfil\epsfbox{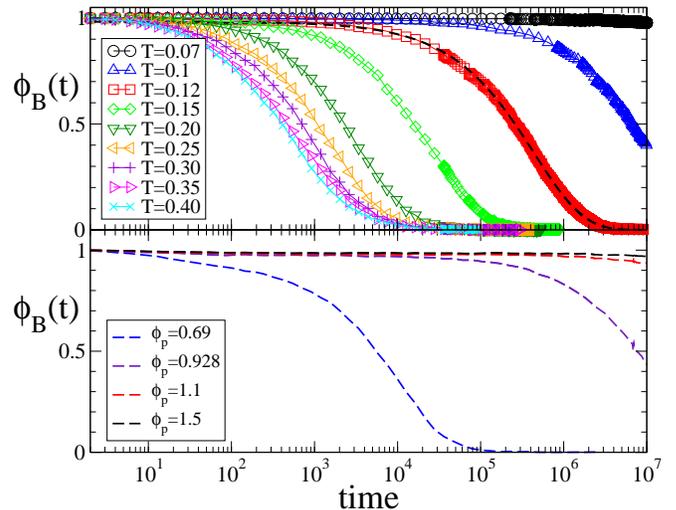}\hfil}
\caption{ Bond correlation function $\phi_B(t)$ for  $\phi_c=0.16$ ($T$-route, top), and for
 $\phi_c=0.15$ ($\phi_p$-route, bottom). The $\phi_B(t)$ shape can be well fitted by a stretched exponential function with stretching exponent $\beta=0.73$ (dashed line superimposed to the $T=0.12$ curve).}
\label{fig:bond-corr}
\vskip 0.7cm
\end{figure}

Figure \ref{fig:bond-corr} shows the evolution of   $\phi_B(t)$  with
$T$ and $\phi_p$. When dynamics slows down, the shape of  $\phi_B(t)$ is preserved at all $T$ or $\phi_p$.  The shape can be modeled  with high accuracy with a 
 stretched exponential function $A exp(-(t/\tau)^{\beta})$, 
with stretching exponent  $\beta \approx 0.73$. 

An estimate of  the average bond lifetime $\tau_B$ can be defined as 
$\tau_B=\tau/\beta \Gamma(1/\beta)$,  where $\tau$ and $\beta$ are calculated via  stretched exponential fits and $\Gamma$ is the Euler Gamma function. 
\begin{figure}
\hbox to\hsize{\epsfxsize=1.0\hsize\hfil\epsfbox{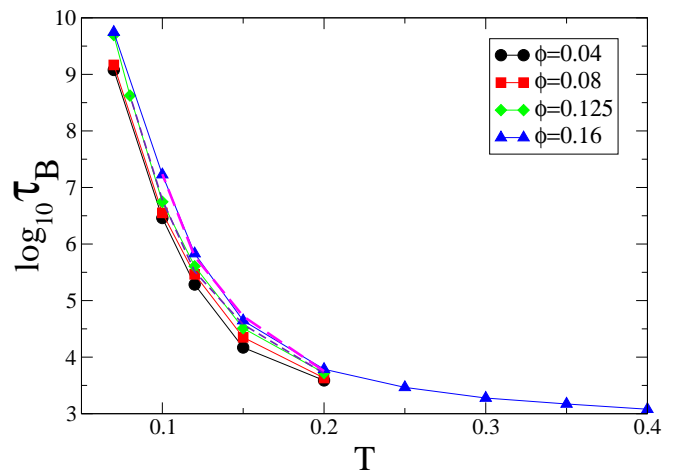}\hfil}
\caption{Temperature dependence of the bond lifetime $\tau_B$ at all
studied densities. 
The short and long dashed lines represent power law fits with exponent 
$\gamma_{\tau}\simeq 3.5$ and $\gamma_{\tau}\simeq4.0$  
and dynamic critical temperatures $T_d=0.084$ and $T_d=0.085$ 
respectively for $\phi_c=0.125$ and  $\phi_c=0.16$.  The $T=0.07$ point, not included in the fits, is shown here only as an indication, since equilibrium is not properly reached at this $T$.}
\label{fig:taubond}
\end{figure}

Fig.\ref{fig:taubond} shows  $\tau_B$ versus $T$. Analogous
considerations to those reported above in discussing the $T$-dependence of $D$
apply. Indeed, $\tau_B(T)$  is consistent with  a power law with
exponent $\gamma_{\tau}$ varying between $3.5$ and $4.0$, 
larger than the one found for $D(T)$, but with consistent predictions for the diverging
$T$.

We  notice that at $T = 0.07$, dynamics is extremely slow and bonds are almost 
unbroken in the time window explored in the simulation. 
It would be interesting to find out if the $T$ dependence of $\tau_B$ crosses 
to a different functional form at low $T$ when all bonds are formed 
and if such cross-over bears some analogies to the cross-over from power-law 
to super Arrhenius observed in glass forming molecular systems.  
Unfortunately, as in the molecular glass cases, the time scale today available to simulation studies does not allow us to resolve this issue. 

\begin{figure}
\hbox to\hsize{\epsfxsize=1.0\hsize\hfil\epsfbox{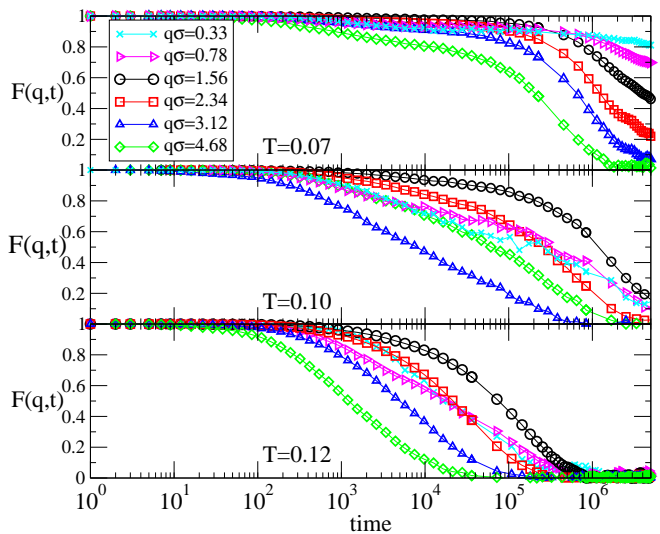}\hfil}
\caption{Wavevector $q$ dependence of the intermediate 
scattering function $F(q,t)$ at $T=0.07,0.10,0.12$ ( from top to bottom ) for
$\phi_c=0.16$. 
The reported  $q\sigma$ values are respectively $0.33, 0.78, 1.56, 2.34, 3.12,  4.68$.}
\label{fig:fqt}
\end{figure}

\begin{figure}
\hbox to\hsize{\epsfxsize=1.0\hsize\hfil\epsfbox{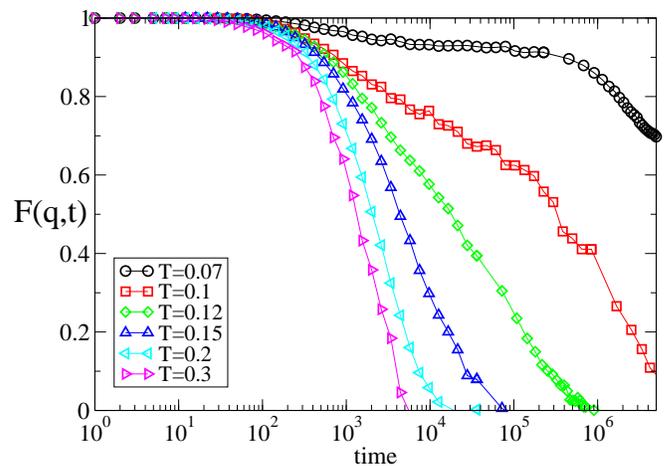}\hfil}
\caption{Temperature dependence of the intermediate scattering function  $F(q,t)$ at $\phi_c=0.16$ and $q\sigma=0.78$. The reported $T$ are $0.07,0.1,0.12,0.15,0.2,0.25,0.3,0.4$.}
\label{fig:fqtT}
\end{figure}

To further compare the arrest observed in the present system and the
slowing down of the dynamics observed in other systems close to
dynamic  arrest, we calculate the collective intermediate scattering
function $F(q,t)$, defined as,
\begin{equation}
F(\vec q,t ) = <\frac{1}{N} \sum_{i,j} 
e^{-i \vec q (\vec r_i(t) -\vec r_j(0))}>
\end{equation}
where the average is calculated  over different starting initial times.
Fig.~\ref{fig:fqt} shows the $q$-dependence of the $F(q,t)$ at three
different $T$ values. The decay of the correlation functions does not
show any appreciable intermediate plateau for any $q$. The functional form of the decay is strongly
dependent on $q$, crossing from an almost $log(t)$ decay at small $q$
to a less stretched decay at large $q$ values.  At the lowest $T$ ($T=0.07$), $F(q,t)$ does not decay to zero any longer, confirming that a non-ergodic state has been reached. 
The non-ergodic behaviour manifests for very
small values of $q\sigma$, in the range of the low $q$  peak in $S(q)$, 
while ergodicity is restored at nearest
neighbor length.
Fig.~\ref{fig:fqtT} contrasts, at fixed $q$ value, the $T$-dependence
of the dynamics.  The shape of $F(q,t)$ is sufficiently different to
conclude that time-temperature superposition does not hold for this observable. 
We also note that at very low $\phi_c$ ($\phi_c=0.04$ or  $0.08$) all density correlation functions decay to zero, within the explored time window, suggesting that cluster diffusion allows for the decay of density fluctuation, even in the presence of a non-ergodic bond restructuring process. This suggests that, 
at low $\phi_c$, density fluctuations are ergodic  in the absence of percolation.

\section{Conclusions:}

In this work we have presented a detailed analysis of the
structural and dynamic properties of a colloidal dispersion in which
the short range attraction is complemented by a screened electrostatic
repulsion.  We have studied one specific choice of the parameters
controlling the repulsive potential. In particular, we have chosen a
screening length comparable to the radius of the colloidal particles. For
this screening length, a study\cite{Mos04a} of the ground state
structure of isolated clusters  showed that the preferential
local structure is composed by a one-dimensional sequence of face-shared
tetrahedra, generating a local six-coordinated structure and a Bernal
spiral shape.

The collective behavior of the system is very much influenced by the
competition between attraction and repulsion, which in the present model sets in when $T$ becomes smaller than 0.2 (in units of the depth of the
attractive part).  The relative location of the
particles, which for $T \gtrsim 0.2$ is  mostly  controlled by translational entropy, 
for $T \lesssim 0.2 $ depends  more and more  on energetic factors.
Between $T=0.2$ and $T=0.1$,  the number of bonded pairs increases significantly , and the local structure evolves progressively toward the six-coordinated one characteristic of the Bernal spiral.  At the lowest 
studied $T$, $T=0.07$, the cluster shape becomes independent of $\phi_c$ and the ground state local configuration becomes dominant. The cluster size distributions at low $T$ show a very clear suppression of clusters of size $ \lesssim 10$, the size requested for the establishment of a bulk component in the spiral configuration. 

Although the majority of particles  tends to preferentially sit in the
6-coordinated configuration, some particles are located in defective regions of the spiral, which act as branching points and favor the formation of large
ramified fractal clusters, whose elementary units are spirals of
finite size.     It is interesting to investigate if the small
energetic cost of branching  allows us to model the spiral segments  as re-normalized monomers.  In support of this possibility, we have detected a 
progressive increase of the cluster fractal dimension for cluster of size $s  \gtrsim100$.  We have also shown that, consistent with the ground state calculations,
clusters of size $s \lesssim 10$ are almost spherical, while
clusters of size $10 \lesssim s \lesssim 100$ are characterized by $d_f \approx 1.25$.

The one-dimensional growth followed by a    
dynamic arrest phenomenon,  observed in this work 
is reminiscent of the aggregation process in several protein 
solution systems~\cite{Nic01c,Pou04a,Ren95a,Manno04}. 
In this class of protein solutions, a variation in the 
external control parameters (temperature, ionic strength, p.H.) 
often trigger  an aggregation process of proteins into 
cylindric clusters which, by branching mechanisms, 
form a macroscopic gel, similarly to what 
takes place in the system here investigated.  
Results reported in this work confirm that, 
as speculated in Ref.~\cite{Mos04a}, there is a range of small but 
finite temperatures in which branching of the one dimensional structure 
is preferred to cluster breaking and that such branching 
does indeed help establishing a connected three dimensional network.

It is important to stress that the dynamic arrest mechanism
observed in this work is very different from the one observed
numerically for the case of $\xi \approx 1.2 \sigma $\cite{Sci04a}. In that case, clusters grow mostly spherical and do not present branching points. The slowing down
of the dynamics in the $\xi \approx 1.2 \sigma$ case  arises from the residual  
repulsive cluster cluster interaction, resulting in the formation of a
cluster phase or a repulsive cluster glass, analogous to the mechanisms suggested for Wigner glass systems.  Indeed, in the arrested state, no percolation was detected.
The arrested state generated via a Wigner glass
transition discussed in Ref.\cite{Sci04a} and the one generated via branching of one-dimensional clusters  discussed in this work, 
although differing only by modest changes in the experimental
conditions, are probably characterized by significantly different
visco-elastic properties. Indeed we expect that the Wigner glass  will be much
weaker than the stiff percolating structure generated by a continuous
sequence of particles tightly bounded to six neighbors.

The system studied in this work is a good candidate for a
thorough comparison with the slowing down characteristic of glass forming
materials.  The numeric "exact" equilibrium particle structure factor
could be used as input in the mode coupling theory, along the lines
theoretically suggested in Ref.\cite{ChenPRE} to provide a full
comparison of the theoretical predictions for the arrest line as well
as for the shape of the correlation functions.  It would be
interesting to quantify the role of the cluster pre-peak in the
structure factor in the predicted slowing down of the dynamics.

Results presented in this work also provide further example of
the existence of equilibrium cluster phases, a phenomenon which is
recently receiving a considerable interest. Cluster phases have
recently been investigated in systems as different as protein
solutions\cite{Strad04,Bagl04,Sedg05}, colloidal
dispersions\cite{Seg01a,Bartlett04,Strad04,Sedg04}, laponite \cite{Ruz04}, liposomic
solutions\cite{Bordi03,Bordi04,Sennato03,Sennato04,Bordi05,Bordi05b},
star-polymers\cite{Stiakpreprint}, aqueous solutions of silver
iodide~\cite{Kegel03}, metal oxides \cite{Kegel04} and in recent
numerical studies\cite{Sci04a,Sator03,Coni04,Kumar05}. In all
cases, the combination of the repulsive interactions with the short
range attraction appears to be crucial in stabilizing the cluster
phase. The high sensitivity of the cluster shape and the final
topology of the arrested state on the detailed balance between range
and amplitude of the attractive and repulsive part of the potential
brought forward by this and previous studies add new
challenges to the modern research in soft condensed matter and to the
possible technological exploitations of these new materials.

A final remark concerns the use of an effective potential,
with state-independent parameters for the description of systems 
in which the screening length can be a function of the colloid 
packing fraction  and in which the  significant changes in 
structure with $T$ (or concentration of depletant) 
may lead to relevant changes in the  cluster surface potential 
or in the spatial distribution of ions.  
The similarity between the numerical data reported in this manuscript and 
the closely related experimental results suggest that, 
despite the approximation adopted in the numerical work, 
the essence of the arrest phenomenon is captured
by the present models.

\section{Acknowledgements}
We thank P. Bartlett and J.  van Duijneveldt for sharing 
their results with us, for discussions and for
calling our attention on the $\hat w$ distributions.
We acknowledge support from MIUR-FIRB and
MRTN-CT-2003-504712.
   
%



%

\bibliographystyle{./apsrev}
\bibliography{./articoli,./altra}

\begin{thebibliography}{47}
\bibitem[{not()}]{notabartlettpot}
\bibinfo{note}{In Ref.\protect\cite{Bartlett04}, the parameter $A$ was
  estimated as $30k_BT$. However, this value is too high, and the potential is
  always positive at the considered $\phi_p$ values. Thus, we chose the value
  of $10$.}
\end{thebibliography}




\end{document}